# Design and hardware evaluation of the optical-link system for the ATLAS Liquid Argon Calorimeter Phase-II Upgrade

**B. Deng,**[a,b,c] **C. Liu,**[c] **D. Gong,**[c] **X. Huang,**[c,d] **S. Hou,**[e] **T. Liu,**[c,1] **H. Sun,**[c,d] **L. Zhang,**[c,d] **W. Zhang,**[c,d] **and J. Ye**[c]

[a] *Department of Electronic Information Engineering, Hubei Polytechnic University, Huangshi, Hubei 435003, P.R. China*

*Zhongyi Environmental Protection College, Yixing Environmental Protection Science and Technology Industrial Park, Yixing, Jiangsu 214200, P.R. China*

[b] *Department of Physics, Southern Methodist University, Dallas, TX 75275, U.S.A*

[c] *Department of Physics, Central China Normal University, Wuhan, Hubei 430079, P.R. China*

[d] *Institute of Physics, Academia Sinica, Nangang, Taipei 11529, Taiwan*

*E-mail*: `tliu@mail.smu.edu`

ABSTRACT: An optical link system is being developed for the ATLAS Liquid Argon Calorimeter Phase-II upgrade. The optical link system is responsible for transmit the data of over 182 thousand detector channels from 1524 Front-End Boards (FEBs) through 26 optical fibers per FEB over 150 meters to the counting room and brings clocks, bunch crossing reset signals and slow control/monitoring signals back to the FEBs. The optical link system is based on the Low-Power GigaBit Transceivers (lpGBTs) and the Versatile optical Transceiver (VTRx+) modules, which both are being developed for the High-Luminosity LHC upgrade. An evaluation board is designed and the major functions of the optical link system are being evaluated. The design of the optical link system and the evaluation of major functions are presented in the paper.

---

[1] Corresponding author.

# 1. Introduction

The readout electronics of the ATLAS Liquid Argon Calorimeter will undergo a major upgrade (Phase-II) for the High-Luminosity Large Hadron Collider (HL-LHC) [1]. In the proposed upgrade, the data of over 182 thousand detector channels are transmitted from the 1524 Front-End boards (FEBs) to the off-detector counting room without any trigger selection [2]. With no trigger selection, the data rate (225 Gbps per FEB) of the on-detector readout electronics after upgrade will be more than 100 times higher than that of the current readout electronics (1.6 Gbps per FEB) [3]. The distance from the detector to the counting room is about 150 meters. Based on the data rate and the distance, optical fiber links are chosen for data transmission. The on-detector electronics operate in a harsh radiation environment. Any component used in the on-detector electronics must tolerate at least a Total Ionizing Dose (TID) of 2.26 kGy, a Nonionizing Energy Loss (NIEL) of $4.9 \times 10^{13}$ 1-MeV-equivalent neutrons/cm$^2$, and a Single-Event Effects (SEE) of $7.7 \times 10^{12}$ hadrons/cm$^2$ (including safety factors), respectively, for a total luminosity of 4000 fb$^{-1}$. To meet the upgrade demands of the ATLAS LAr Calorimeter, a radiation-tolerant optical link system is being developed and a printed circuit board (PCB) called the link demonstrator board has been designed. The major functions of the optical link system are being evaluated with one or two demonstrator boards. In this paper, we present the design of the optical link system and the evaluation of major functions.

# 2. Overall architecture of the optical link system and the demonstrator system

Figure 1 is the block diagram of the overall architecture of a FEB. In the figure, 128 analog channels are amplified and shaped in 32 4-channel Pre-Amplifiers (PAs, including shaping amplifiers) [4] and then are digitized in 32 8-channel Analog-to-Digital Converters (ADCs) [5]. Each analog channel is digitized with two different gains simultaneously in order to achieve a 16-bit dynamic range. The digitized data are transmitted from the detector to the off-detector counting room through optical links. These uplinks are called data links. We also need control links to provide clock and control/monitoring support. Each FEB has 22 simplex data links and 2 duplex control links. Optical links are based on lpGBTs [6] and VTRx+ modules [7], which both are being developed for the High-Luminosity LHC upgrade. In this paper, the lpGBTs responsible for data links will be called data lpGBTs, and the lpGBTs responsible for control links will be called control lpGBTs.

Figure 2(a) is the block diagram of the link demonstrator system with two demonstrator boards and corresponding FPGAs. The lpGBT on the first demonstrator board (Demo1) operates in the transceiver mode and is responsible for control links, while the lpGBT on the second demonstrator (Demo2) operates in the transmitter mode and is responsible for the data link. Two FPGA development kits (BE-FPGA1 and BE-FPGA2, Part No. KC705 produced by Xilinx) emulate the off-detector FPGAs [8] and two more FPGA development kits (FE-FPGA1 and FE-FPGA2) emulate ADC data generation and check downlink data. The lpGBT on Demo1 recovers a clock for the lpGBT on Demo2. A clock generator (Part number SI5338EVB from Silicon Laboratory) provides the clocks for the two off-detector FPGAs. A photograph of the setup is shown in Figure 2(b).

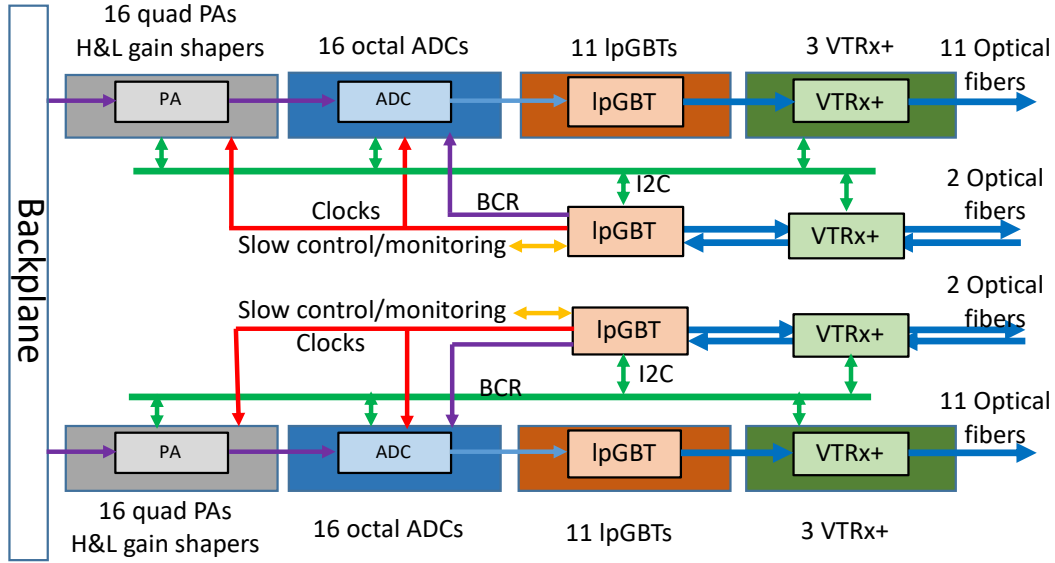

Figure 1. Block diagram of the overall architecture of FEB.

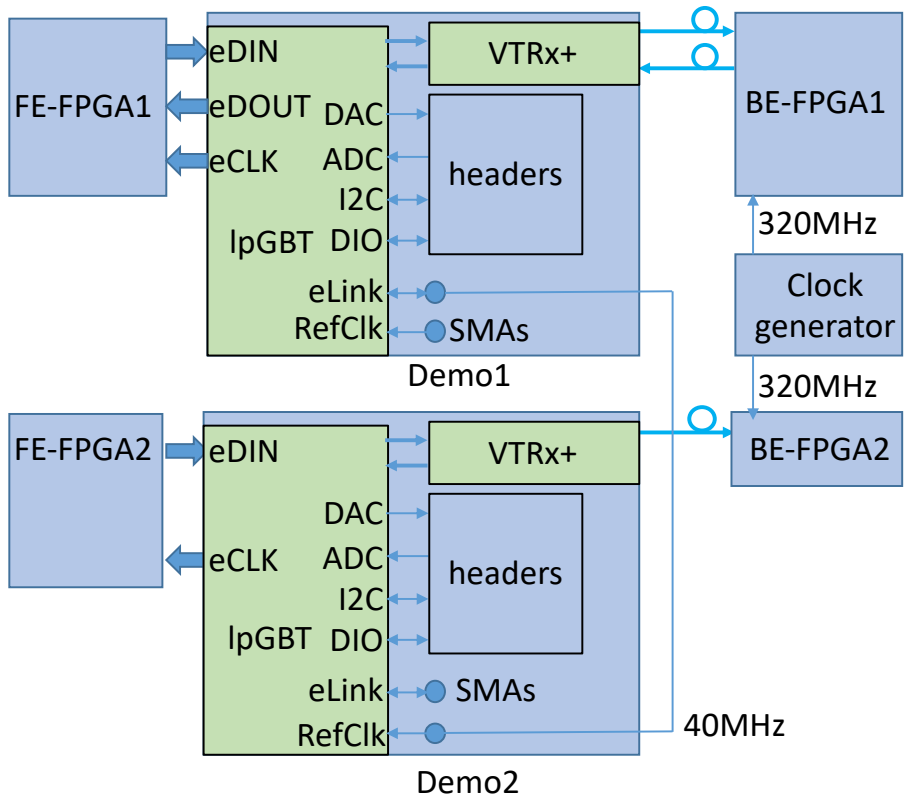

(a)

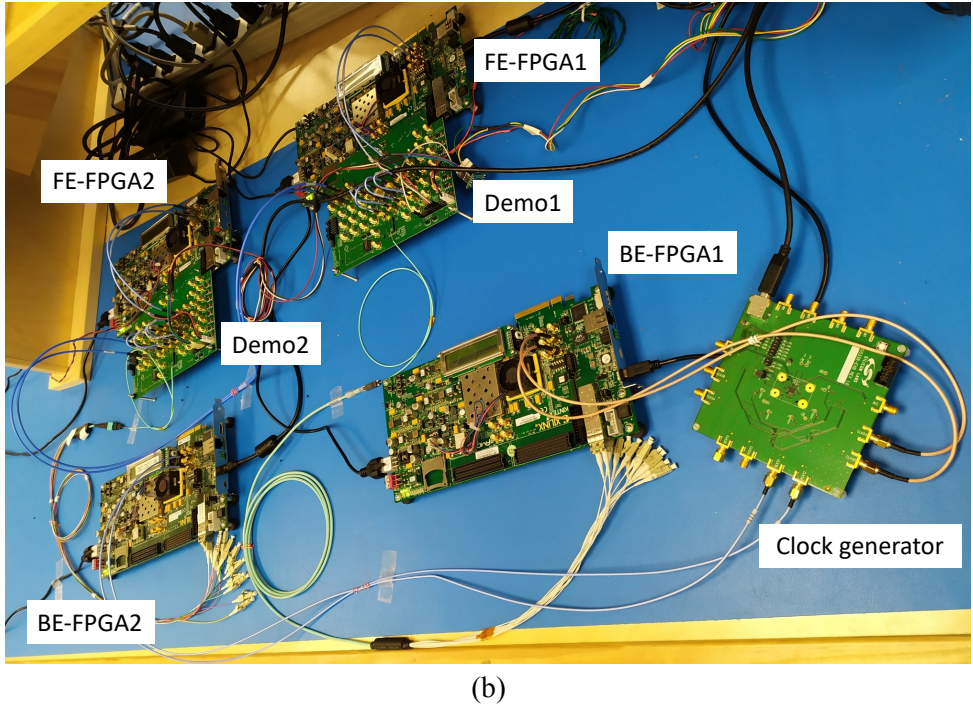

(b)

**Figure 2.** Block diagram (a) and photograph (b) of the optical link demonstrator system.

## 3. Design and evaluation of data transmission

### 3.1 Design of data transmission

The interface from ADCs to lpGBTs is shown in Figure 3(a). Each ADC takes 8 analog signals, half in low gain and the other half in high gain, from the preamplifiers. The digitized signals of ADCs are serialized into 640 Mbps and output to lpGBTs in the format of CERN-Specified Low-Voltage Signaling (CLVS) format [9]. It should be noted that lpGBTs have intrinsic bit alignment mechanics to ensure that lpGBTs do not sample the data on bit boundaries [10]. However, word boundaries are lost after serialization. Therefore, we must introduce work boundary recovery mechanics. In addition to the eight normal data output channels, each ADC has two extra output channels call the frame channels. The word in a frame channel is defined as {1, PARITY, FIRSTEVENT, BCID[4:0], 00000000}. PARITY is the parity of the frame word. FIRSTEVENT and BCID[4:0] will be discussed in Section 5. We use the frame channel to recover the word boundary and the full BCID count in the off-detector FPGA.

Each lpGBT acquires the data from 14 electrical ports (e-ports) at 640 Mbps, encodes the data by adding a frame header, an Internal Control (IC) field, an External Control (EC) field, and Forward Error Correction (FEC) code, and serializes the encoded data into a serial sequence at 10.24 Gbps. As shown in Figure 3(a), the frame channel and the 8 normal data channels of the first ADC go to the first lpGBT. The first half of the channels (a frame channel and 4 normal data channels) of the second ADC go to the first lpGBT and the second half go to the second lpGBT. The frame channel and the 8 normal data channels of the third ADC go to the second lpGBT. Both the first ADC and the third ADC each have an unused frame channel (frame[1] in Figure 3(a)). The trace length of all serial data outputs from an ADC to the corresponding lpGBT should be kept as similar as possible.

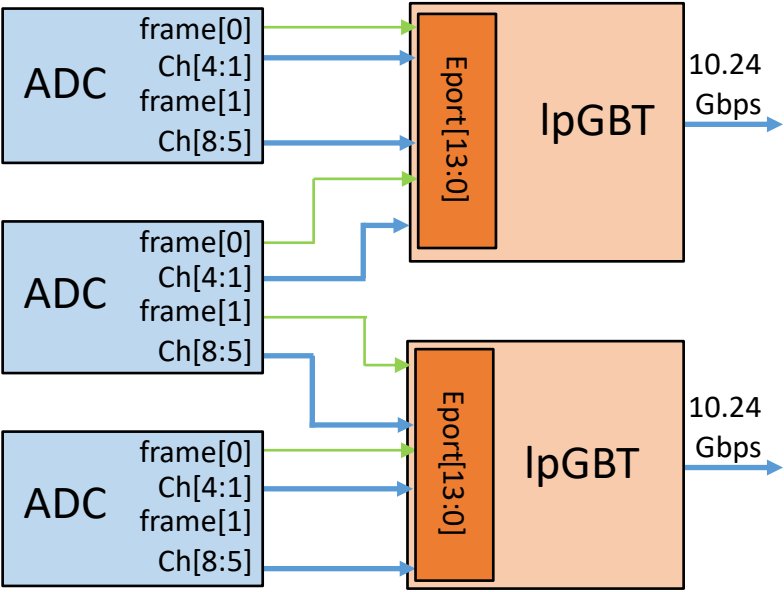

(a)

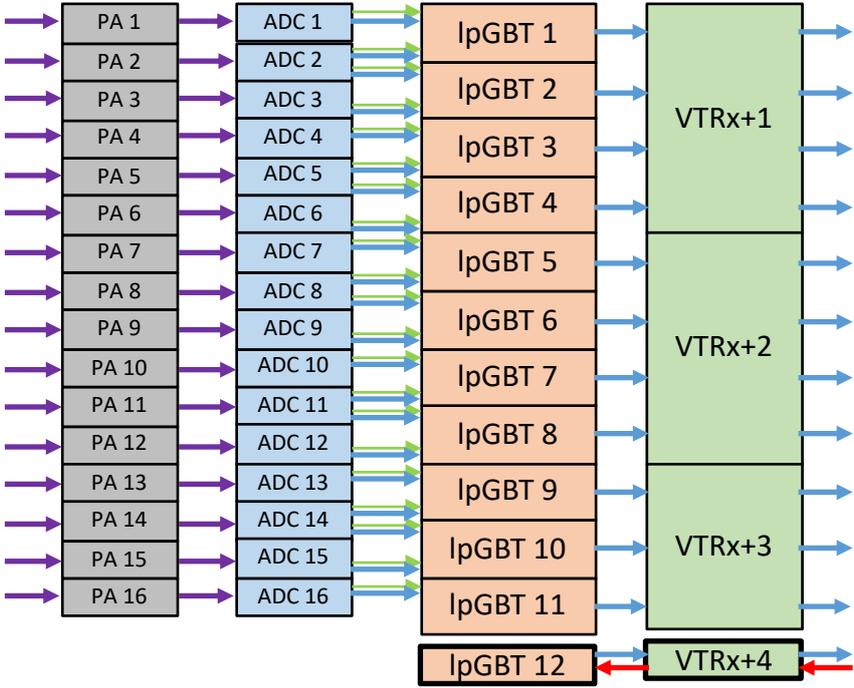

(b)

Figure 3. Mapping from ADCs to lpGBTs (a) and data flow of ½ FEB (b).

Figure 3(b) is the data flow of half FEB. In the figure, lpGBTs 1-11 are for data links and lpGBT 12 is for control links. Each half FEB has 16 preamplifiers, 16 ADCs, 12 lpGBTs, 4 VTRx+ modules, and 13 optical fibers.

### 3.2 Evaluation of data transmission

The data transmission has been evaluated using two demonstrator boards, as shown in Figure 2. Figure 4 displays the eye diagrams of the downlink at 2.56 Gbps and the uplink at 10.24 Gbps after passing the VTRx+ module.

During the test, the uplink data were generated in FE-FPGA2. The frame channels frame[1:0] followed the definition described in the previous section. The normal data channels Ch[8:5] and Ch[4:1] were each either a 4-bit binary counter or pseudo-random binary sequence with preset initial values. The word boundary was recovered correctly in BE-FPGA2. The uplink data were checked in BE-FPGA2 using a logic analyzer tool without any error for more than 15 minutes. The bit error rate (BER) is estimated to be $1\times10^{-12}$, the BER specification of lpGBT/VTRx+ and the ATLAS liquid argon calorimeter, with a confidence level of 99.99%.

The downlink data, either four 4-bit binary counters or pseudo-random binary sequences with preset initial values, were generated in BE-FPGA1. The data were received without any error for more than 15 minutes, corresponding to the BER of $1\times10^{-12}$ with a confidence level of 90.0%.

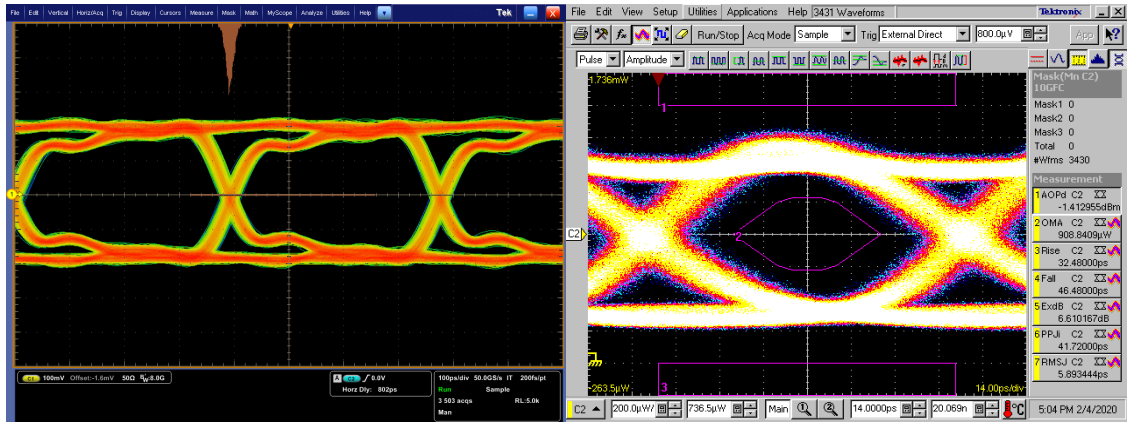

(a)                  (b)

**Figure 4.** Eye diagrams of the 2.56 Gbps downlink (a) and the 10.24 Gbps uplink (b) after passing through VTRx+ modules.

## 4. Design and evaluation of Clock distribution

Each preamplifier needs a 40 MHz clock to generate an internal clock for its Inter-Integrated Circuit ($I^2C$) functional part, while each ADC requires a 40 MHz clock for sampling and a 640 MHz clock for output data serialization. Figure 5 is the block diagram of clock distribution on a half FEB. The lpGBT 12 operating in the transceiver mode recovers 11 40

MHz clocks from the received downlink data for lpGBTs 1-11. Each data lpGBT operating in the transmitter includes a Phase-Locked Loop (PLL) and regenerates 40 MHz and 640 MHz clocks for preamplifiers and ADCs. The clocks for the preamplifiers are phase fixed, while the clocks for ADCs are phase adjustable.

The clock distribution scheme has been evaluated. With two demonstrator boards as shown in Figure 2, we observed a recovered 40 MHz clock from the lpGBT on Demo1 and the regenerated 40 MHz and 640 MHz clocks on Demo2. All recovered and regenerated clocks are locked to the off-detector clock. The random jitter of the regenerated 40 MHz clocks is about 2.2 ps (RMS). In our application, the 40 MHz clock is used in ADCs as the sampling clock and sensitive to jitter. The measured random jitter meets the system requirements of 5 ps (RMS). Significant periodic jitter components of about 23 ps (peak-to-peak) are observed in the regenerated 640 MHz clock. The periodic jitter components are to be reduced in the next version of the lpGBT. We confirmed that the phase adjustable clock can be tuned with a step as low as 48.8 ps.

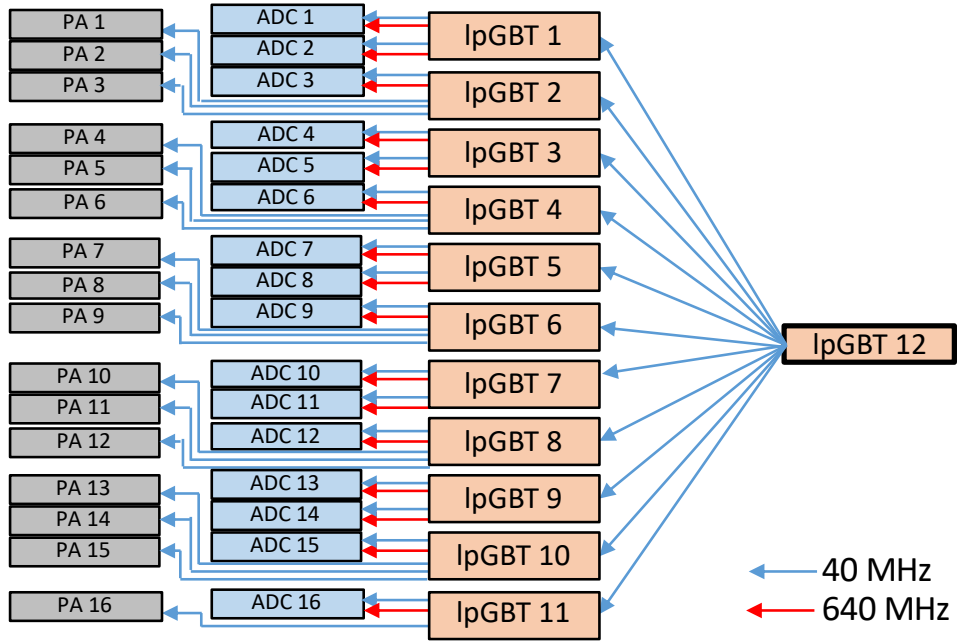

**Figure 5.** Block diagram of clock distribution.

## 5. Design and evaluation of the BCR distribution

Bunch Crossing Identification (BCID) is a 12-binary-bit count from 0 to 3563 used to align the detector data with the LHC proton bunches, which collide at a frequency of 40 MHz. Each ADC has an internal BCID counter. Limited by the bandwidth and the frame definition, only a five-bit count is transmitted out of each ADC. As a compensation, and a single-bit FIRSTEVENT flag, which is 1 in the LHC clock cycle of 25 ns immediately after a BCR signal is asserted and 0 for the rest of the time, is included in the frame channel. The off-detector

FPGA recovers the 12-bit BCID count based on the five-bit count and the FIRSTEVENT flag. To operate the FIRSTEVENT flag and the BCID counter, each ADC requires a BCR signal to set its FIRSTEVENT flag and to reset its internal counter.

The BCR distribution scheme is shown in Figure 6(a). Each control lpGBT has 16 transmitter e-ports and each e-port provides a BCR signal for an ADC. The data rate of each e-port is 80 Mbps. By changing the data transmitted out of the e-port, the phase of the BCR signal can be adjusted in steps of 12.5 ns.

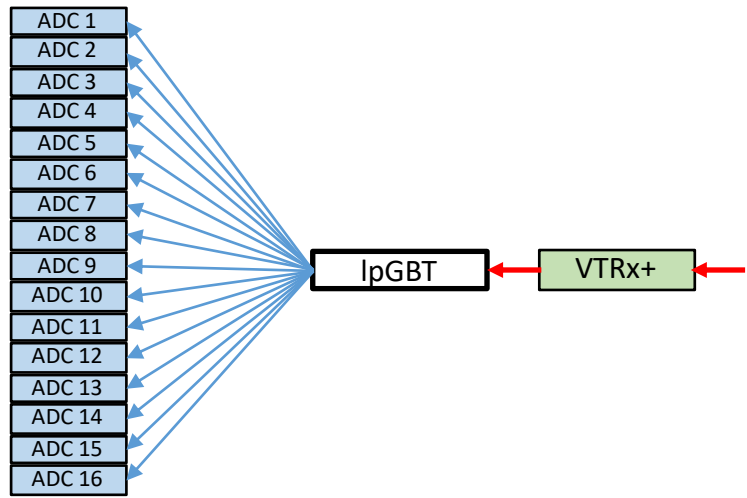

(a)

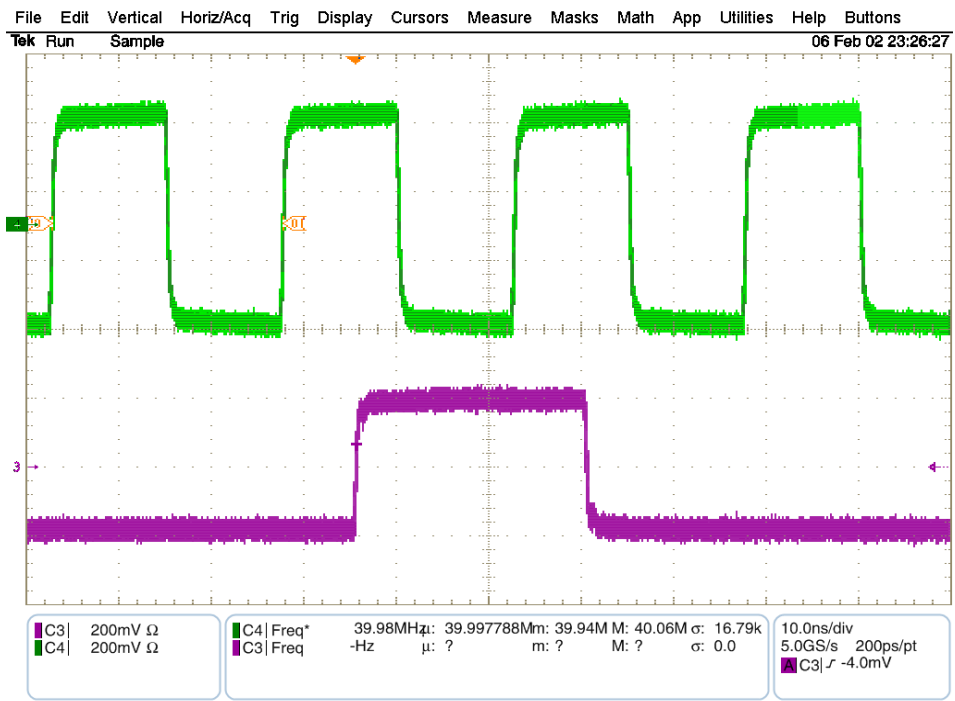

(b)

**Figure** 6. Block diagram of BCR distribution (a) and measured waveform of the BCR signal (b).

We have evaluated a BCR distribution scheme on the link demonstrator board. Figure 6(b) displays the measured waveform of a BCR signal in purple related to a 40 MHz clock as the trigger in green. The pulse width of the measured BCR signal is 25 ns and the period is 89.1 µs, or 3564 cycles of the LHC bunch crossing clock. We also confirm that the phase of the BCR signal can be shifted in steps of 12.5 ns.

## 6. Design and evaluation of the ASIC configuration

The preamplifiers and the ADCs installed on the FEBs need to be configured remotely. Though lpGBTs and VTRx+ modules have intrinsic electrical fuses (e-fuses), which provide non-volatile configuration, it is still preferable to have remote configuration capability. Preamplifiers, ADCs, lpGBTs, and VTRx+ all have an $I^2C$ interface. Figure 7 is the block diagram of the ASIC configuration. Each control lpGBT, which has three $I^2C$ controllers (M0, M1, and M2), is responsible for half FEB. The $I^2C$ controllers M0 and M1 each configure eight preamplifiers and eight ADCs. The lpGBT 12 can be configured via its IC channel. The data lpGBTs 1-10 can share $I^2C$ controllers with preamplifiers and ADCs. The three $I^2C$ controllers each configure a VTRx+ module. VTRx+ 3 has to be configured through the lpGBT 11 via the EC channel.

The function of the ASIC configuration has been evaluated. We verified that the lpGBT on Demo1 configured the VTRx+ on the same board and the lpGBT on Demo2 correctly. We confirmed that the lpGBT on Demo1 can be controlled remotely through its IC channel. The verification of an lpGBT configured via the EC channel is still ongoing.

## 7. Design and evaluation of other slow control/monitoring

In addition to the $I^2C$ configuration of ASICs, control links are responsible for other slow control and monitoring. The block diagram of slow control/monitoring is shown in Figure 8. First, we can turn on or off all power supply modules and reset ASICs. This capability is necessary in case a device is stuck and needs a power cycle or a reset. This is implemented by using Digital Input and Output (DIO) ports of a control lpGBT. The on-board pull-up or pull-down resistors of each DIO ensure a default setting and correspondingly the normal operation of the board during the initialization of control lpGBT. It should be noted that a control lpGBT can only power cycle or reset the control lpGBT that is located on the other half board. Second, we monitor the temperatures at various positions. This is implemented via the embedded ADCs of the control lpGBTs. Third, we monitor all power supply voltages via the embedded ADC of lpGBTs. Since each lpGBT can monitor only 8 voltages, there are not enough control lpGBTs to monitor all power supply voltages. We use data and control lpGBTs to monitor the power supply voltages. The input signal range of the embedded ADC of the lpGBT is from 0 V to 1 V. When a power supply voltage to be monitored is larger than 1 V (for example, 2.5 V for preamplifiers), a resistor-division network must be used.

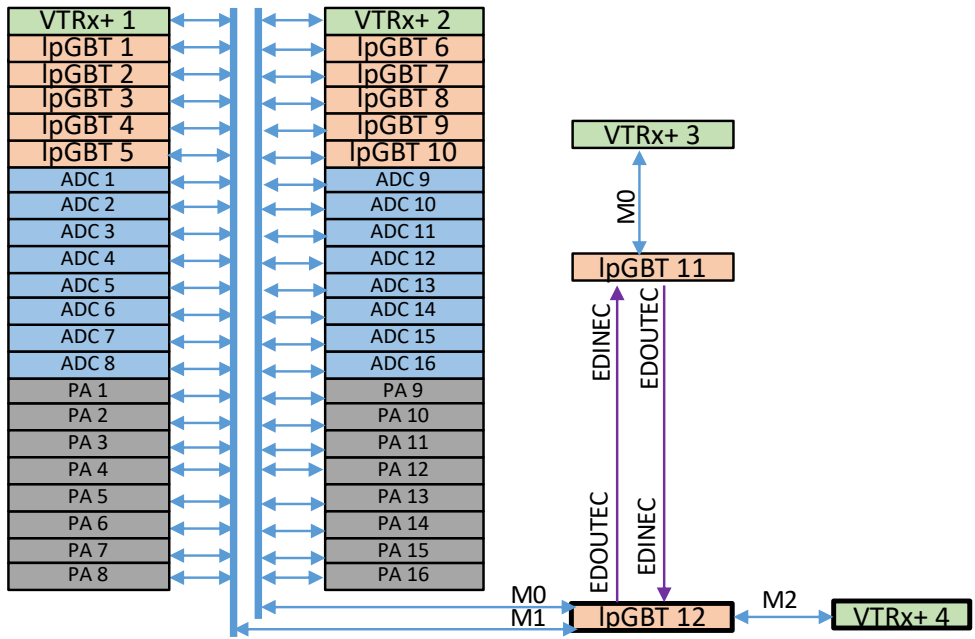

**Figure 7.** Block diagram of I²C configuration.

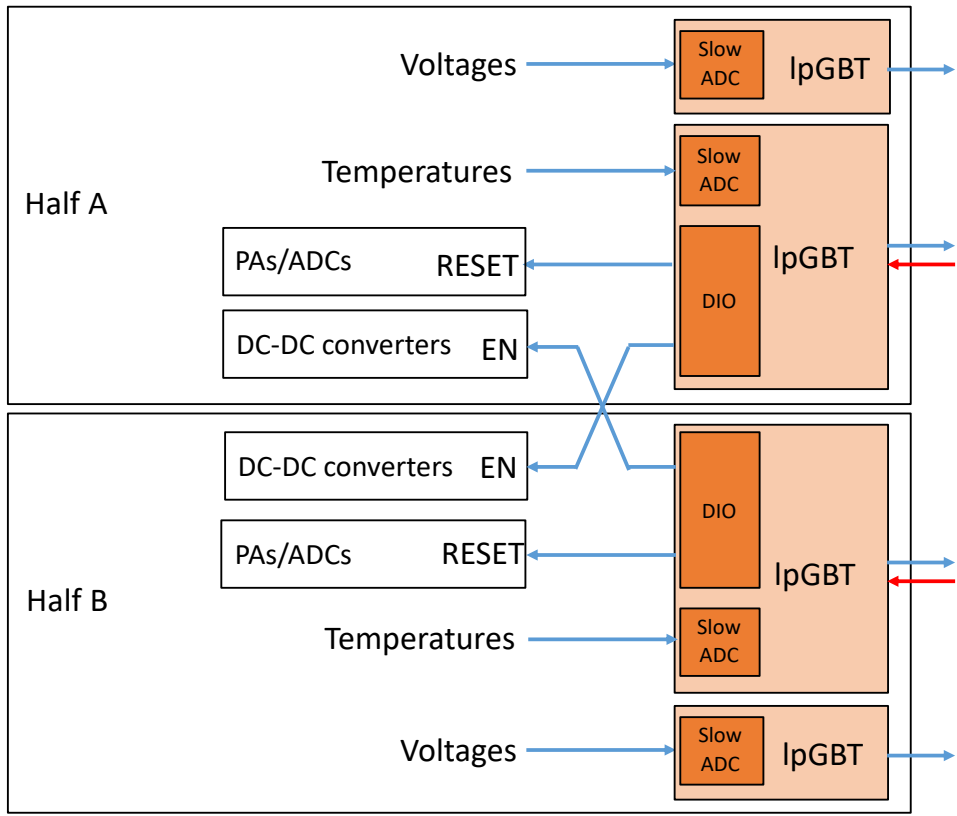

**Figure 8.** Block diagram of other control/monitoring.

The slow control/monitoring scheme other than the I²C configuration has been evaluated on the demonstrator board. For digital input, we adjusted the input voltage of the DIO pin and read back the digital state in software. For digital output, we preset a low or high state in software and measured the output voltage. Particularly, the DIO of the lpGBT on Demo1 reset the lpGBT on Demo2 correctly. The lpGBT monitored the room temperature and voltages correctly. For voltage monitoring, the input voltage was generated using a programmable power supply and a resistor division network. Though the conversion time of the embedded ADC is less than 1 μs, the rate to monitor temperatures and voltages is at the order of milliseconds because the conversion start and the poll of conversion completion are controlled in software.

## 8. Conclusion

An optical link system is being produced for the ATLAS LAr Calorimeter Phase-II upgrade. A link demonstrator board is designed and major functions of the optical link system have been evaluated, though extensive tests are still ongoing.

A vertical slice of FEB called the prototype link board, which includes 24 lpGBTs and 8 VTRx+ modules, will be designed. The prototype link board will be tested with the same FPGAs. We will use the prototype link board to study the redundancy design of the control links. No redundancy is considered in the link demonstrator board except the crossover power cycle and reset of control lpGBTs. Redundant distribution of clocks, BCRs, and I2C configuration will be investigated in the prototype link board. The prototype link board will also address the system-level issues such as trace/fiber routing scheme, mechanical support, power dissipation, and cooling. Take the power dissipation as an example. The power dissipation of the optical link system (24 lpGBTs and 8 VTRx+ modules) on every FEB is estimated to be up to 20 W. If we assume the conversion efficiency of 85% in DC-DC converters and a voltage drop of 1 V in Low Drop Out (LDO) regulators, the optical link system on each FEB consumes about 42 W, over 1/3 of the power consumption budget (110 W) of the whole FEB. The possibility of eliminating the LDO regulators and minimizing the noise of DC-DC converters with a multiple-phase synchronous operation will be explored in the prototype link board.

**Acknowledgments**

We acknowledge the support of the US-ATLAS R&D grant for the ATLAS LAr Phase-II upgrade project. We are grateful for the support from Drs. Szymon Kulis, Paulo Moreira, David Porret, Jan Troska, Francois Vasey, and other lpGBT and Versatile Link+ members at CERN. We thank Drs. Christophe de LA Taille and Sylvie Blin at OMEGA/IN2P3/Ecole Polytechnique/CNRS, John Parsons, Gustaaf Brooijmans, and Jaro Ban at Columbia University, Timothy Andeen at the University from Texas at Austin, Nicolas Dumont Dayot at LAPP, Bernard Dinkespiler at CPPM, and Arno Straessner at Technische Universität Dresden for beneficial discussions.

## References

[1] ATLAS Collaboration, Technical Design Report for the Phase-II Upgrade of the ATLAS LAr Calorimeter, CERN-LHCC-2017-018, CERN, Geneva, 2017.

[2] T. Berger-Hryn'ova, Development of the ATLAS Liquid Argon (LAr) Calorimeter Readout Electronics for the HL-LHC, presented at CHEF2019 - Calorimetry for the High Energy Frontier 2019, 28 November 2019, Proceedings available: ATL-COM-LARG-2020-002.

[3] N.J. Buchanan, L Chen, D.M. Gingrich, S Liu, H Chen, D Damazio, et al, ATLAS liquid argon calorimeter front end electronics, 2008 *JINST* 3 P09003.

[4] C. De La Taille et al., Performance of LAUROC1, ATLAS Liquid Argon Upgrade ReadOut Chip, presented at TWEPP 2019 Topical Workshop on Electronics for Particle Physics, Santiago de Compostela, Spain, September 3, 2019.

[5] T. Andeen, Development of the ATLAS Liquid Argon Calorimeter Readout Electronics for the HL-LHC, presented at the 18[th] International Conference on Calorimeter in Particle Physics (CALOR 2018), Eugene, Oregon, USA, May 25, 2018.

[6] P. Moreira et al., The lpGBT: a radiation tolerant ASIC for Data, Timing, Trigger and Control Applications in HL-LHC, presented at TWEPP 2019 Topical Workshop on Electronics for Particle Physics, Santiago de Compostela, Spain, September 3, 2019.

[7] J. Troska et al., *The VTRx+, an Optical Link Module for Data Transmission at HL-LHC*, PoS (TWEPP-17) 048.

[8] J. Mendez et al., *New LpGBT-FPGA IP: Simulation model and first implementation*, PoS (TWEPP2018) 059.

[9] D. Guo et al., The eTx line driver and the eRx line receiver: two building blocks for data and clock transmission using the CLPS standard, presented at TWEPP 2019 Topical Workshop on Electronics for Particle Physics, Santiago de Compostela, Spain, September 3, 2019.

[10] S. Kulis et al., A multi-channel multi-data rate circuit for phase alignment of data in the lpGBT, presented at TWEPP 2019 Topical Workshop on Electronics for Particle Physics, Santiago de Compostela, Spain, September 6, 2019.